\documentclass[conference]{IEEEtran}
\IEEEoverridecommandlockouts
\usepackage{cite} 
\usepackage{amsmath,amssymb,amsfonts,bbm} 
\usepackage{algorithmic}
\usepackage{tabularx} 
\usepackage{amsmath, bm} 
\usepackage{amsmath, autobreak} 
\usepackage[ruled, linesnumbered]{algorithm2e} 
\usepackage{subcaption}
\usepackage{soul} 
\usepackage{balance} 
\usepackage{color} 
\usepackage{float} 
\usepackage{autobreak} 
\usepackage{graphicx} 
\usepackage{textcomp} 
\usepackage{xcolor} 
\usepackage{booktabs} 
\usepackage{multirow} 
\usepackage{bbm}
\usepackage{dsfont} 
\usepackage{footnote} 
\usepackage[]{mathtools} 
\usepackage{lipsum} 
\makesavenoteenv{algorithm} 
\usepackage{diagbox}
\usepackage{lineno}
\usepackage{tikz}
\usepackage{makecell}
\usepackage{amssymb}
\usepackage{mathrsfs}
\usepackage{calligra}
\usepackage[numbers,sort&compress]{natbib}
\usepackage[left=1.62cm,right=1.62cm,top=1.9cm]{geometry}

\newtheorem{remark}{Remark}

\definecolor{gblue}{rgb}{0, 0.3, 0.59}
\definecolor{purple}{rgb}{0.8, 0.2, 1}

\allowdisplaybreaks[4]

\begin{document}

\title{
Sample-Efficient Misconfiguration Classification for Network Resilience in Wireless Communications}

\author{
\IEEEauthorblockN{
Xin~Hao, 
Massimo~Piccardi,
Chenhan~Zhang,
Vijaya~Durga~Chemalamarri,
Qiwen~Jiang,\\
Wei~Ni,~\IEEEmembership{Fellow,~IEEE},
and~Raymond~Owen}
\IEEEauthorblockA{Faculty of Engineering and Information Technology, University of Technology Sydney, NSW, Australia\\
E-mails: \{xin.hao, massimo.piccardi, chenhan.zhang, vijayadurga.chemalamarri, jean.jiang, wei.ni, ray.owen\}@uts.edu.au}

}

\markboth{Journal of \LaTeX\ Class Files,~Vol.~14, No.~8, August~2015} %
{Shell \MakeLowercase{\textit{et al.}}: Bare Demo of IEEEtran.cls for IEEE Journals}

\maketitle

\begin{abstract}
As modern wireless communication networks grow increasingly complex, network outages driven by the inconsistency between dynamic topologies and protocol configurations have become a critical concern. To solve this issue, we mathematically formulate a protocol misconfiguration classification problem as a graph-based learning task and solve it with our proposed EtaGATv2 algorithm, an edge-type-aware graph attention network with dynamic attention. EtaGATv2 addresses two critical challenges: i) it captures non-uniform symptom propagation for protocol misconfiguration classification tasks, where certain network paths and nodes become critical for diagnosis, and ii) it extracts protocol-specific features from heterogeneous routing protocols with distinct message-passing behaviors by utilizing edge-type-aware transformations. Experiments across diverse and real-world topologies demonstrate that EtaGATv2 reaches state-of-the-art performance with $50\%$ of the training samples, making it particularly suitable for networks with dynamic topologies and limited negative-labeled data.
\end{abstract}

\begin{IEEEkeywords}
Wireless communications, network resilience, graph attention networks, misconfiguration classification
\end{IEEEkeywords}

\IEEEpeerreviewmaketitle

\section{Introduction}\label{sec_introduction}
In recent years, wireless communication network outages have occurred worldwide with alarming frequency~\cite{Survey_contemporary_resilience_2025, Xin_GSID}. These outages make network resilience, i.e., the ability to provide services despite network degradations and failures~\cite{definition_resilience_D2R2}, increasingly critical. Much research attention has been devoted to restoring network outages caused by network component failures, such as links and entities~\cite{ComMag_detect_before_outage_occur, inet_link_failure_IoT_mesh, Xin_BCDRL}. However, these approaches fall short as misconfiguration has become the dominant cause of network outages~\cite{oecd2025_misconf, Ray_Sovereign_Functions}, especially with the widespread adoption of template-based configurations in modern wireless networks~\cite{cisco_crosswork_template, cisco_base_configuration}. This is because template-based misconfiguration propagates to multiple devices, transforming the problem from pinpointing individual failed components to classifying which parameter causes the outage, a challenge that is critically severe under dynamic network topologies but lacks effective solutions.

The first challenge is that the misconfiguration classification task in wireless communication networks exhibits distinct locality characteristics. When template-based misconfiguration occurs, diagnostic symptoms propagate non-uniformly, with certain paths and nodes becoming critical for identifying the misconfigured parameter. This demands adaptive mechanisms to selectively focus on how diagnostic information concentrates through specific network regions at varying granularities. Although existing graph-based approaches demonstrate strong capabilities for tasks in dynamic wireless communication networks, including topology construction~\cite{inet_topology_construction}, routing optimization~\cite{RouteNet}, and protocol configuration synthesis~\cite{NeurIPS_BGP}. However, none of these works incorporates mechanisms for such selective emphasis on critical local patterns of protocol misconfigurations.

The second challenge arises from the heterogeneous nature of routing protocols in wireless communication networks, as different routing protocols exhibit distinct message-passing behaviors and state update mechanisms~\cite{TNSE_routing_dynamic}. This protocol-specific heterogeneity demands feature extraction that can capture these distinct behaviors, yet conventional graph neural network approaches treat protocol types uniformly during message passing in the graphs~\cite{Xin_HML, Xin_ICC2023}. This challenge is further intensified by the dynamic nature of wireless communication networks~\cite{liuchang_2023predictive}. Therefore, it remains to be addressed how to differentiate protocol-specific propagation patterns during message passing, which is essential for accurate protocol misconfiguration classification.


In this paper, we propose an edge-type-aware GATv2 (EtaGATv2) algorithm to enable sample-efficient misconfiguration classification in networks with topologies. Our main contributions are summarized as:
\begin{itemize}
    \item We mathematically formulate the template-based misconfiguration classification as a graph-based learning problem, capturing template-wide propagation across multiple devices rather than isolated device failures, thereby enabling generalization to diverse network topologies.
    \item We design the EtaGATv2 algorithm that employs dynamic attention mechanisms to capture locality-based symptom propagation at varying granularities, and edge-type-aware transformations to extract protocol-specific features, enabling selective focus on critical diagnostic information in resilient wireless communication networks.
    %
    \item Experimental results validate that EtaGATv2 achieves superior convergence with significantly lower label dependency. By effectively leveraging the structural priors of wireless communication networks, our approach addresses the long-standing challenge of learning from sparse negative samples, reaching state-of-the-art performance with $50\%$ of the training samples.
\end{itemize} 


\section{System Model}

\begin{figure}[t]
    \centering
    \includegraphics[width=0.9\linewidth]{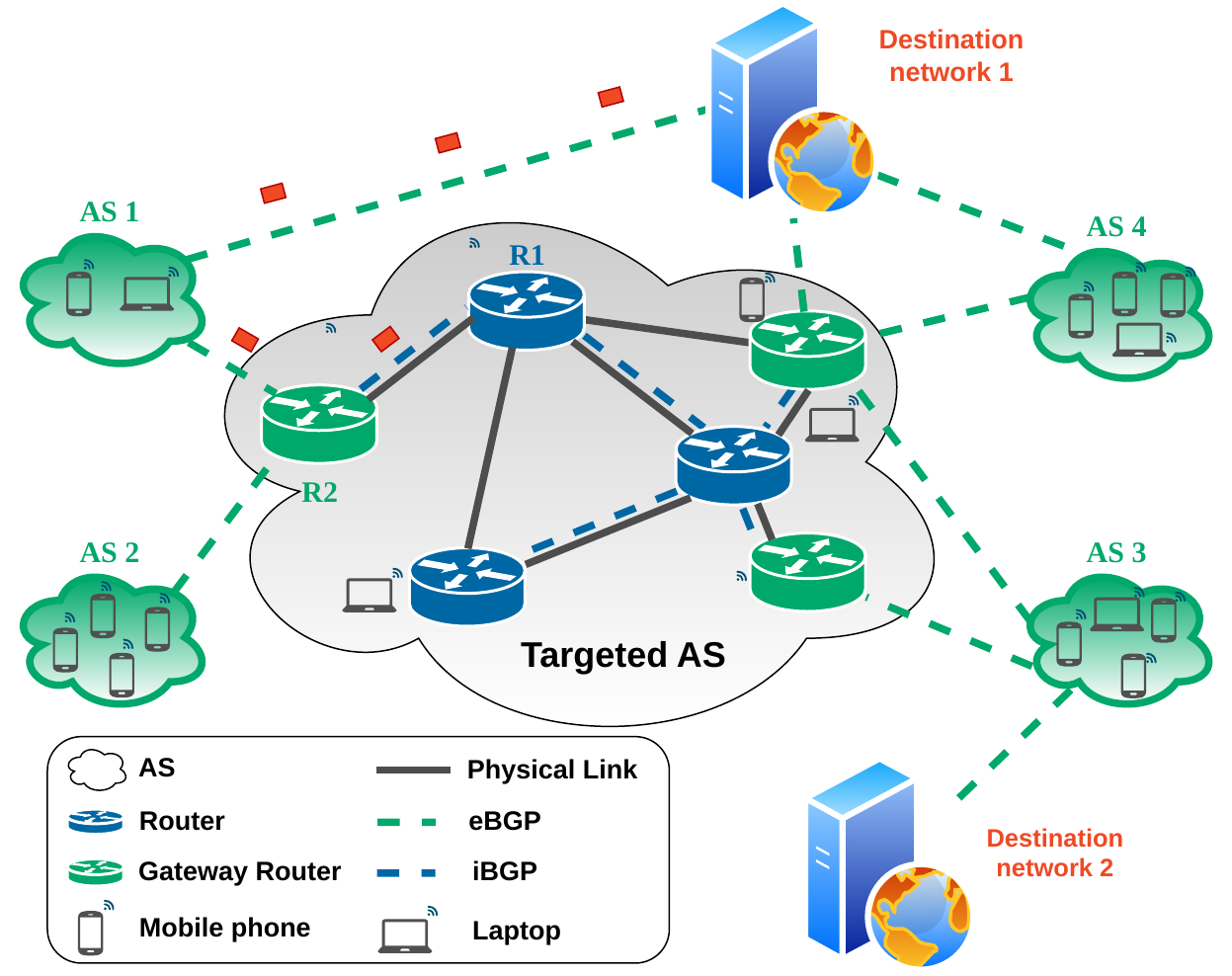}
    \caption{According to the current protocol configurations, the data packets (red blocks) from router R1 to destination network 1 are forwarded through R2 along the OSPF-computed path (solid line), using the route learned via iBGP from R2, then advertised via eBGP from AS1.}
    \vspace{-10pt}
    \label{fig_system_model}
\end{figure}

As shown in Fig.~\ref{fig_system_model}, in the considered network scenario, a targeted AS routes traffic to multiple destination networks, and is managed by an Internet Service Provider (ISP). The AS consists of internal routers interconnected by physical links, with some gateway routers connecting to external ASes that serve heterogeneous wireless devices, e.g., mobile phones and laptops. The gateway routers exchange routing information with neighboring ASes via eBGP, enabling end-to-end connectivity for wireless users across different domains.

Two routing protocols govern the network behavior. The Open Shortest Path First (OSPF) protocol computes the shortest paths between internal routers based on link weights. The Border Gateway Protocol (BGP) comprises external BGP (eBGP), which exchanges routing information with external ASes, and internal BGP (iBGP), which distributes this external routing information among internal routers.

The ISP uses template-based configurations to manage protocol parameters across multiple devices. When a configuration template contains a misconfiguration, this error propagates to all devices using that template, causing network-wide routing anomalies. Our goal is to classify which protocol parameter template is misconfigured based on observed routing symptoms.



\subsection{Graph-based Network Model}\label{sec_network_model}
We first model this network as a graph, denoted by $\mathcal{G} = (\mathcal{V}, \mathcal{E})$, where the node and edge sets are given by
\begin{align}
\mathcal{V} = \mathcal{V}_\mathrm{router} \cup \mathcal{V}_\mathrm{dst} \cup \mathcal{V}_\mathrm{exas}
\text{~~and~~}
\mathcal{E} = \mathcal{E}_\mathrm{inter} \cup \mathcal{E}_\mathrm{intra},
\end{align}
respectively, where 
$\mathcal{V}_\mathrm{router}=\{ v_{\mathrm{router}_1}, \cdots, v_{\mathrm{router}_I}\}$, 
$\mathcal{V}_\mathrm{dst}   =\{ v_{\mathrm{dst}_1},    \cdots, v_{\mathrm{dst}_K}   \}$, and 
$\mathcal{V}_\mathrm{exas}  =\{ v_{\mathrm{exas}_1},   \cdots, v_{\mathrm{exas}_M}  \}$ 
represent the sets of the internal routers, the destination networks, and the external ASes, respectively; while $\mathcal{E}_\mathrm{inter}$ and $\mathcal{E}_\mathrm{intra}$ denote the connections between internal-internal nodes and internal-external nodes, respectively.

To further leverage the protocol-specific propagation features, we classify the edges into different types according to the networking protocols. Specifically, the type set is denoted as $\mathcal{T} = \{\tau_\mathrm{OSPF}, \tau_\mathrm{eBGP}, \tau_\mathrm{iBGP}, \cdots \}$, where OSPF edges, $\mathcal{E}_{\text{OSPF}}$, carry weighted intra-domain routing information, eBGP edges, $\mathcal{E}_{\text{eBGP}}$, establish inter-domain peering, and iBGP edges, $\mathcal{E}_{\text{iBGP}}$, distribute internal BGP routes.

\subsection{Protocol Configuration and Misconfiguration Models}
The routing behavior of the network is determined by protocol configurations, which specify parameter values for OSPF and BGP. In modern network management, operators use configuration templates to deploy consistent parameter settings across multiple devices~\cite{cisco_crosswork_template}. For example, a single OSPF link weight template may instantiate the same weight value on dozens of links. A misconfiguration occurs when such a template contains an incorrect parameter value. Unlike isolated device failures, where a single component malfunctions, template-based misconfigurations cause the same error to propagate to all devices using that template, resulting in correlated routing anomalies across multiple network locations. This propagation pattern creates distinct diagnostic symptoms that enable classification of the misconfigured parameter.

Following~\cite{NeurIPS_BGP}, the configuration is defined as
\begin{equation}
\mathcal{C} = \mathcal{C}_{\mathrm{BGP}} \cup \mathcal{C}_{\mathrm{OSPF}},
\end{equation}
where $\mathcal{C}_{\mathrm{BGP}}$ and $\mathcal{C}_{\mathrm{OSPF}}$ denote the BGP and OSPF configuration sets, respectively, and 
\begin{equation}
\mathcal{C}_{\mathrm{BGP}} 
= 
\left\{
\theta_{m,k}^{(1)}, \cdots, \theta_{m,k}^{(n)}, \cdots, \theta_{m,k}^{(N)}
\right\},
\end{equation}
where $\theta_{m,k}^{(n)}$ represents the $n$-th BGP attribute for BGP route advertisement that is from the $m$-th external AS $v_{\mathrm{exas}_m}$ to the $k$-th destination network $v_{\mathrm{dst}_k}$. These attributes control the BGP route selection according to the BGP decision process, and can be local preference, AS-path length, etc. 
%
The OSPF configuration set is given by
\begin{equation}
\mathcal{C}_{\mathrm{OSPF}} = \{\phi_{ij} \mid (v_{\mathrm{router}_i}, v_{\mathrm{router}_j}) \in \mathcal{E}_{\mathrm{inter}}\},
\end{equation}
where $\phi_{ij} \in [1, \phi_{\max}]$ and $\phi_{ij} \in \mathbb{Z}^+$ is the weight for the internal link from the $i$-th router to the $j$-th router. 

We consider template-based protocol configurations~\cite{cisco_crosswork_template}, where a single configuration template instantiates parameters across multiple network devices. We categorize template-level misconfigurations based on the affected protocol parameters
\begin{equation}
\mathcal{F} = \mathcal{F}_{\mathrm{OSPF}} \cup \mathcal{F}_{\mathrm{BGP}} \cup \{f_0\},
\end{equation}
where set $\mathcal{F}_{\mathrm{OSPF}} = \{f_{\phi}\}$ represents misconfigurations in the OSPF link weight template affecting the $\phi_{ij}$ assignments, set $\mathcal{F}_{\mathrm{BGP}} = \{f_{\theta^{(1)}},  \ldots, f_{\theta^{(N)}}\}$ represents misconfigurations in the BGP attribute templates, where $f_{\theta^{(n)}}$ corresponds to errors in the $n$-th dimension of $\boldsymbol{\theta}_{m,k}$, and $\{f_0\}$ represents the misconfiguration-free scenario.


\subsection{Specification Model}~\label{sec_specification_model}
Given the configuration set $\mathcal{C}$ that determines network behavior, the ISP needs to verify whether the actual routing outcomes align with their intended behaviors. To formalize this verification process, we introduce specifications that represent the desired routing behaviors. Our specification set comprises forwarding, reachability, and isolation, which is given by
\begin{equation}
\mathcal{S} = \mathcal{S}_{\mathrm{fwd}} \cup \mathcal{S}_{\mathrm{reach}} \cup \mathcal{S}_{\mathrm{iso}},
\end{equation}
where $\mathcal{S}_{\mathrm{fwd}} = \{s_{\mathrm{fwd}(i,k,j)} = (v_{\mathrm{router}_i}, v_{\mathrm{dst}_k}, v_{\mathrm{router}_j})\}$; $\mathcal{S}_{\mathrm{reach}}=\{s_{\mathrm{reach}(i,k,j)}=(v_{\mathrm{router}_i}, v_{\mathrm{dst}_k}, v_{\mathrm{router}_j})\}$; and $\mathcal{S}_{\mathrm{iso}}=\{s_{\mathrm{iso}(i,j,k,m)}=(v_{\mathrm{router}_i}, v_{\mathrm{router}_j},v_{\mathrm{dst}_k}, v_{\mathrm{dst}_{m}})\}$. 
More specifically, $s_{\mathrm{fwd}(i,k,j)}$ specifies that the $i$-th router forwards traffic to the $k$-th network via the $j$-th router; $s_{\mathrm{reach}(i,k,j)}$ specifies that the traffic targeted for the $k$-th network through the $i$-th router also passes through the $j$-th router; and $s_{\mathrm{iso}(i,j,k,m)}$ specifies that in the link between the $i$-th and $j$-th routers, the traffic destined for the $k$-th and $m$-th networks is isolated.


\section{Proposed EtaGATv2 Solution}\label{sec_Proposed_methodology}
\subsection{Problem Formulation}
Given a network $\mathcal{G}$, with ISP intended specifications $\mathcal{S}$ and observed configurations $\tilde{\mathcal{C}}$, our objective is to identify the most likely misconfiguration class from the template-level misconfiguration set $\mathcal{F}$. This problem is formulated as
\begin{equation}
\label{eq_formulated_problem_raw}
\mathcal{P}1:
~~~~
\hat{f} = \arg\max_{f \in \mathcal{F}} p(f \mid \mathcal{G}, \mathcal{S}, \tilde{\mathcal{C}}).
\end{equation}
To solve problem $\mathcal{P}1$, we first calculate the observed specifications $\tilde{\mathcal{S}}$ from the observed configurations $\tilde{\mathcal{C}}$ by executing the protocols by using $\tilde{\mathcal{S}}=\textsc{Prot}(\tilde{\mathcal{C}}, \mathcal{G})$~\cite{NeurIPS_BGP}. Next, we check the existence of misconfigurations by $f_\mathrm{check}(\tilde{\mathcal{S}}) = \mathds{1}\{\tilde{\mathcal{S}} \ne \mathcal{S}\}$.
If $f_\mathrm{check}(\tilde{\mathcal{S}})=0$, no misconfiguration exists, and the specification check process terminates. Otherwise, when $f_\mathrm{check}(\tilde{\mathcal{S}})=1$, we proceed to the misconfiguration classification stage.

We note that this specification-level check desirably excludes cases where the configuration deviations do not violate specifications, thereby improving classification accuracy.
If misconfiguration exists, problem $\mathcal{P}1$ is reformulated as a learning process. We use a neural network with parameter vector $\mathbf{\omega}$ to map its input to a probability distribution over possible misconfigurations. The most likely misconfiguration class can be obtained by solving 
\begin{equation}
    \mathcal{P}2:~~~~
    \hat{f} = \arg\max_{f \in \mathcal{F}\backslash\{f_0\}} p_{\omega}(f \mid \mathcal{G}, \mathcal{S}, \hat{\mathcal{C}}),
\end{equation}
where $\hat{\mathcal{C}}$ denotes the configuration set that has been confirmed to contain misconfigurations, and $p_{\omega}(\cdot)$ represents the predicted probability of misconfigurations, which is computed by
\begin{equation}
    p_{\omega}(f \mid \mathcal{G}, \mathcal{S}, \hat{\mathcal{C}}) 
    = \texttt{softmax}(\texttt{MLP}(z)),
    \label{eq_probability_of_classfified_misconfiguration}
\end{equation}
where $z$ is the graph-level readout.



\subsection{Design Specifics}\label{sec_EtaGATv2_alg}
The design specifics include two key aspects: a \textit{dynamic attention mechanism} that adapts to individual node misconfiguration characteristics, and an \textit{edge-type-aware scheme} to obtain protocol-specific propagation features.

\subsubsection{Dynamic Attention Mechanism}
For reference, we recall the attention coefficients between nodes $u$ and $v$ calculated using GAT~\cite{GAT_ICLR_2018} and GATv2~\cite{GATv2}, on which EtaGATv2 is based, as 
\begin{equation}
    e_{uv}^\mathrm{GAT} = \texttt{LeakyReLU}(\mathbf{a}^\top \cdot [\mathbf{W}\mathbf{h}_u \| \mathbf{W}\mathbf{h}_v]),
\end{equation}
and
\begin{equation}
    e_{uv}^\mathrm{GATv2} = \mathbf{a}^\top \cdot \texttt{LeakyReLU}(\mathbf{W} \cdot [\mathbf{h}_u \| \mathbf{h}_v]),
\end{equation}
respectively, where $\mathbf{a}$ is a learnable attention parameter vector, $\mathbf{W}$ is a learnable weight matrix for linear transformation, $\mathbf{h}_u$ and $\mathbf{h}_v$ represent the input feature vectors of nodes $u$ and $v$, respectively, and $\|$ denotes the concatenation operation.
GATv2's key improvement over GAT is that the attention vector, $\mathbf{a}$, is placed after the non-linearity, preventing $\mathbf{a}$ and $\mathbf{W}$ from collapsing into a single parameter, and creating a ``dynamic'' attention mechanism that scores the source and target nodes jointly rather than separately. 

\begin{remark}
    This dynamic attention mechanism is particularly valuable for protocol misconfiguration classification for dynamic wireless communication networks, where the relevance of neighboring nodes varies significantly across misconfiguration classes. For example, for a misconfiguration causing link failures, route re-convergence alarms from downstream routers become highly relevant for diagnosis, while nodes on unaffected paths should receive minimal attention despite their connectivity.
\end{remark}

\subsubsection{Edge-Type-Aware Scheme} 
In wireless communication networks, different protocols exhibit distinct alarm propagation patterns. For example, OSPF weight misconfigurations propagate through link-state updates along $\mathcal{E}_{\text{OSPF}}$ edges, affecting shortest-path computations. However, BGP local preference errors spread via $\mathcal{E}_{\text{iBGP}}$ sessions, altering route selection decisions. 
These protocol-specific propagation behaviors motivate our edge-type-aware design, where each edge type $\tau \in \mathcal{T}$ is associated with dedicated parameters $\mathbf{a}_\tau$ and $\mathbf{W}_{\tau}$. 

\begin{remark}
The edge-type-aware scheme enables the GAT to learn distinct propagation patterns for different protocols, providing a stronger inductive bias that reduces the feature space and required training samples.
\end{remark}

\subsubsection{EtaGATv2}\label{section_EtaGATv2}
Our EtaGATv2 integrates the dynamic attention and edge-type-aware mechanism, thereby the attention coefficient of EtaGATv2 is given by
\begin{equation}
    e_{uv,\tau}^\mathrm{EtaGATv2} 
    = \mathbf{a}_\tau^\top \cdot \texttt{LeakyReLU}(\mathbf{W}_{\tau} \cdot [\mathbf{h}_u \| \mathbf{h}_v]),
\end{equation}
where $\mathbf{a}_\tau$ and $\mathbf{W}_{\tau}$ are learnable edge-type-aware attention vector and weight matrix, respectively.


%
The attention coefficient is then normalized using \texttt{softmax} to obtain the attention weight
\begin{equation} 
\label{eq_attention_score} 
    \alpha_{uv,\tau}^\mathrm{EtaGATv2} 
    = \texttt{softmax} \left( e_{uv,\tau}^\mathrm{EtaGATv2} \right). 
\end{equation}
In the message-passing process, the $v$-th node aggregates information from its neighbors according to their edge types through type-specific message passing
\begin{equation} 
\label{eq_EtaGATv2_message_passing} 
m_{v}^\mathrm{EtaGATv2} 
= \sum_{\tau \in \mathcal{T}} \sum_{u \in \mathcal{N}_\tau(v)} \alpha_{uv,\tau}^\mathrm{EtaGATv2}  \cdot \mathbf{W}_{\tau} \mathbf{h}_u, 
\end{equation}
where $\mathcal{N}_\tau(v)$ denotes the neighbor set connected to the $v$-th node by edge type $\tau$.

\begin{remark}[]
In EtaGATv2, the dynamic attention captures how identical misconfigurations exhibit different characteristics from heterogeneous nodes, while edge-type awareness extracts protocol-specific propagation features. 
\end{remark}

The overall training procedure performs $L$ layers of message passing, aggregates graph-level embeddings via pooling, and classifies misconfigurations by multi-layer perceptrons (MLPs).
%
%
The step-by-step training procedure (Algorithm~\ref{alg_EtaGATv2}) embeds node features (line 4), injects random misconfigurations (line 5), performs $L$ layers of message passing (lines 6--8), aggregates graph-level embeddings via mean pooling (line 10), and classifies misconfigurations by MLPs (line~11).


\begin{algorithm}[t]
\algsetup{linenosize=\small} \small 
\SetAlgoLined
\caption{The proposed EtaGATv2 Algorithm}
\label{alg_EtaGATv2}
\KwIn{
    Training dataset $\mathcal{D} = \{G_1, G_2, ..., G_N\}$ where each $G_i = (V_i, E_i, X_i, T_{E_i})$, with $X_i$ and $T_{E}$ represent the router states and edge types, respectively;
    Number of misconfiguration classes $|\mathcal{C}|$;
    Learning rate $\eta$, training epochs $T$, hidden dimension $h$;
    Number of attention heads $H$, and number of GNN layers $L$.
}
\KwOut{Well-trained parameters of the neural network $\omega^*$.}
\textit{Initialize}: Model parameters $\omega$, including node embedding encoder, dynamic edge-type-aware GATv2 layers, and misconfiguration classifier.
\For{$t = 1 \cdots T$}
{
  \For{each batch $\mathcal{B}$ from $\mathcal{D}$}
  {
    Add bidirectional edges and self-loops.
    
    Initialize node embedding: $H^{(0)} = \texttt{Emb}(X) \in \mathbb{R}^{|V| \times h}$.
    
    Randomly sample and inject the misconfigurations $\mathbf{c} = [c_1, \dots, c_{|\mathcal{B}|}] $ for graphs $\{G_i\}_{i=1}^{|\mathcal{B}|}$ in the batch.
    
    \For{$\ell = 1 \cdots L$}{
      Apply GATv2-based dynamic edge-type-aware message passing by $m_v^{(\ell)} = \sum_{\tau \in \mathcal{T}} \sum_{u \in \mathcal{N}_\tau(v)} \alpha_{uv}^{(\ell)} \cdot \mathbf{W}_{\tau} \mathbf{h}_u^{(\ell-1)}$.
      
      Update node embeddings: $h_v^{(\ell)}$.
    }
    
    Calculate graph-embedding: 
        $z = \frac{1}{|V|} \sum\nolimits_{v \in V} h_v^{(L)}$.
    
    Get probability of misconfiguration class by eq.~\eqref{eq_probability_of_classfified_misconfiguration}.
    
    Apply loss function and stochastic gradient descent.
  }
}
\end{algorithm}
%

\subsection{Computational Complexity Analysis}
\subsubsection{Rule-based Algorithm}\label{sec_RB_alg}
To contextualize the computational complexity of EtaGATv2, we first analyze a rule-based (RB) approach as a theoretical reference. The core intuition is symptom-fault mapping. Different misconfiguration types produce distinct patterns of specification violations, which can be matched against predefined fault signatures derived from domain knowledge.
Algorithm~\ref{alg_RB} operationalizes this through three steps. First (line 2), we identify violated specifications as $\Delta\mathcal{S} = \mathcal{S} \setminus \tilde{\mathcal{S}}$. Second (line 3), we decompose $\Delta\mathcal{S}$ into forwarding, reachability, and isolation violations. Third (lines 5-10), we match this pattern against predefined fault signatures.

The matching process (line 6) uses symptom pattern weights $w_f: \mathcal{F} \times \mathcal{O} \to [0,1]$ that encode domain knowledge. For instance, OSPF weight faults $f_{\phi}$ typically alter shortest paths, yielding high $w_{f_{\phi}}^{\mathrm{fwd}}$, while BGP attribute faults $f_{\theta^{(n)}}$ primarily affect route selection, yielding high $w_{f_{\theta^{(n)}}}^{\mathrm{reach}}$. The algorithm computes a weighted matching score for each fault type and outputs the one with the highest score as $\hat{f}$ (lines 7-10).

RB decomposes violated specifications (lines 2-3) with complexity $O(|\Delta\mathcal{S}|)$, then computes matching scores for all $|\mathcal{F}|$ fault classes (lines 5-10) with complexity $O(|\mathcal{F}| \cdot |\Delta\mathcal{S}|)$, where $\Delta\mathcal{S} = \mathcal{S} \setminus \tilde{\mathcal{S}}$ denotes violated specifications. From the specification model in Section~\ref{sec_specification_model}, $|\mathcal{S}| = |\mathcal{S}_{\mathrm{fwd}}| + |\mathcal{S}_{\mathrm{reach}}| + |\mathcal{S}_{\mathrm{iso}}|$, where $|\mathcal{S}_{\mathrm{fwd}}| = O(|\mathcal{V}_{\mathrm{router}}|^2 \cdot |\mathcal{V}_{\mathrm{dst}}|)$, $|\mathcal{S}_{\mathrm{reach}}| = O(|\mathcal{V}_{\mathrm{router}}|^2 \cdot |\mathcal{V}_{\mathrm{dst}}|)$, and $|\mathcal{S}_{\mathrm{iso}}| = O(|\mathcal{V}_{\mathrm{router}}|^2 \cdot |\mathcal{V}_{\mathrm{dst}}|^2)$. Thus, $|\Delta\mathcal{S}| \leq |\mathcal{S}| = O(|\mathcal{V}_{\mathrm{router}}|^2 \cdot |\mathcal{V}_{\mathrm{dst}}|^2)$. Therefore, the computational complexity of RB is given by
\begin{equation}
O_{\mathrm{RB}}^{\mathrm{inf}} 
= O(|\mathcal{F}| \cdot |\Delta\mathcal{S}|)
= O(|\mathcal{F}| \cdot |\mathcal{V}_{\mathrm{router}}|^2 \cdot |\mathcal{V}_{\mathrm{dst}}|^2).
\label{eq_complexity_RB}
\end{equation}


\begin{algorithm}[t]
\algsetup{linenosize=\small} \small 
\SetAlgoLined
\caption{The Rule-based (RB) Algorithm}
\label{alg_RB}

\KwIn{
Observed configuration $\tilde{\mathcal{C}}$; Observed specification $\tilde{\mathcal{S}}$;
Expected specification $\mathcal{S}$;
Fault class set $\mathcal{F} \backslash \{f_0\}$.
}

\KwOut{Diagnosed fault type $\hat{f}$.}

\textit{Initialise}: $\texttt{score}_{\max}=0$ and $\hat{f} \gets \text{null}$.

Identify violated specification: $\Delta\mathcal{S} \gets \mathcal{S} \setminus \tilde{S}$.

Decompose violations: $\Delta\mathcal{S}_{\mathrm{fwd}} \gets \Delta\mathcal{S} \cap \mathcal{S}_{\mathrm{fwd}}$, 
$\Delta\mathcal{S}_{\mathrm{reach}} \gets \Delta\mathcal{S} \cap \mathcal{S}_{\mathrm{reach}}$,
$\Delta\mathcal{S}_{\mathrm{iso}} \gets \Delta\mathcal{S} \cap \mathcal{S}_{\mathrm{iso}}$.

Define symptom pattern weight $w_f: \mathcal{F} \times \mathcal{O}
\to [0,1]$ by domain knowledge, where $ \mathcal{O} = \{\mathrm{fwd},\, \mathrm{reach},\, \mathrm{iso}\}$.

\For{each fault type $f \in \mathcal{F} \backslash \{f_0\}$}{
    
    Compute matching score:
    $\texttt{score}(f) = 
    {\sum_{o \in \mathcal{O}} w_f^{o} \, \big|\Delta\mathcal{S}_{o}\big|}/{\big|\Delta\mathcal{S}\big|}$,
    
    \If{$\texttt{score}(f) > \texttt{score}_{\max}$}{
        $\texttt{score}_{\max} \gets \texttt{score}(f)$.
        
        $\hat{f} \gets f$.
    }
}


\end{algorithm}

\subsubsection{EtaGATv2}
EtaGATv2 performs $L$ layers of message passing, and at each layer, the edge-type-aware mechanism over $|\mathcal{E}|$ edges with $H$ heads and dimension $h$ costs $O(|\mathcal{E}| \cdot H \cdot h^2)$, while the node updates for $|\mathcal{V}|$ nodes cost $O(|\mathcal{V}| \cdot h^2)$. The graph readout and classification add a negligible $O(|\mathcal{V}| \cdot h + h \cdot |\mathcal{F}|)$. Thus, the overall complexity is
\begin{equation}
O^{\mathrm{EtaGATv2}} = O\left(L \cdot (|\mathcal{E}| \cdot H \cdot h^2 + |\mathcal{V}| \cdot h^2)\right).
\label{eq_complexity_EtaGATv2}
\end{equation}
Eq.~\eqref{eq_complexity_EtaGATv2} shows that EtaGATv2 exhibits linear complexity in the network scale, $|\mathcal{V}|$, which is a highly desirable property. 
%

\subsubsection{Comparative Analysis}
Comparing eqs.~\eqref{eq_complexity_RB} and \eqref{eq_complexity_EtaGATv2}, RB exhibits quadratic complexity $O(|\mathcal{V}_{\mathrm{router}}|^2 \cdot |\mathcal{V}_{\mathrm{dst}}|^2)$ while EtaGATv2 achieves linear complexity $O(|\mathcal{E}| + |\mathcal{V}|)$. More critically, adapting to topology changes, RB requires manual weight redesign by domain experts, whose effectiveness is inherently bounded by human knowledge and difficult to scale. In contrast, EtaGATv2 enables automated fine-tuning without human intervention, making it particularly suitable for dynamic wireless communication networks where topology changes are frequent and labeled misconfiguration data are scarce.

\begin{table}[t]
\caption{Key Parameters of Different Datasets}
\label{tab_dataset_configurations}
\centering
\setlength{\tabcolsep}{3pt}
\renewcommand{\arraystretch}{1.3}
{
\begin{tabular}{l|rrr}
\toprule 
\toprule 
\diagbox[width=10em]{\textbf{Parameters}}{\textbf{Networks}}  & \textbf{Baseline} & \textbf{~~Larger-Scale} & \textbf{Real-World}
\vspace{2pt}
\\
\hline
Topology Type                        & Synthetic         & Synthetic             & ~~~Topology Zoo 
\vspace{2pt}
\\
\hline
Router Numbers                  & 16--23            & 24--31                & By topologies \\
Dest. Network Numbers           & 4--7              & 10--15                & 4--7 \\
Gateway Node Numbers            & 3                 & 7--9                  & 3 \\
\texttt{fwd} Querie Numbers     & 8--12             & 25--35                & 8--12 \\
\texttt{reach} Querie Numbers   & 4--7              & 15--20                & 4--7 \\
\texttt{iso} Querie Numbers     & 10--30            & 10--30                 & 10--30 
\vspace{2pt}
\\
\hline
Train Sample Numbers            & 1024              & --                    & -- \\
Test Sample Numbers             & 100               & 100                   & 100 \\
\bottomrule
\bottomrule
\end{tabular}
}
\end{table}

\begin{figure}[t]
\centering
\begin{subfigure}[t]{0.46\linewidth}
    \centering
    \includegraphics[width=4.4cm, height=3.5cm]{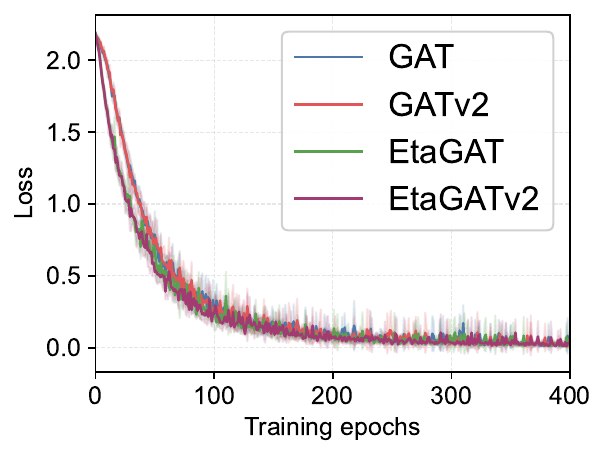}
    \subcaption{Average loss}
    \label{fig1_training_loss_4algo}
\end{subfigure}
\begin{subfigure}[t]{0.46\linewidth}
    \centering
    \includegraphics[width=4.4cm, height=3.5cm]{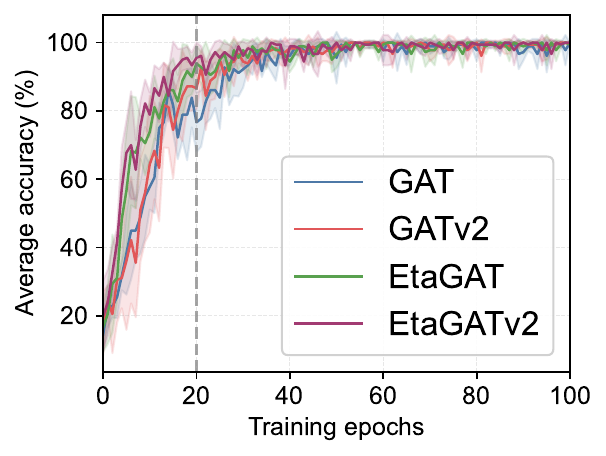}
    \subcaption{Average accuracy}
    \label{fig2_training_accuracy_4algo_moving_avg}
\end{subfigure}
\caption{Training results of multiple independent runs using the baseline dataset given in Table.~\ref{tab_dataset_configurations}, {with 1024 samples in each training epoch}.}
\label{fig_sim_training}
\end{figure}



\section{Performance Evaluation and Analysis}

%
\subsection{Experimental Setup}
Following~\cite{NeurIPS_BGP}, we evaluate our model on network instances generated using a BGP/OSPF protocol simulator, with each network instance having seven template-based configuration parameters: one OSPF link weight (maximum value 32) and six BGP route attributes (local preference, multi-exit discriminator, origin of routing, AS path length, Cisco's local router preference, and index of the external AS). For each network instance, we inject a single template-level misconfiguration by randomly selecting one misconfiguration class from $\{f_1, f_2, \ldots, f_7\}$, where $f_1$ corresponds to OSPF link weights and $f_2$--$f_7$ correspond to the six BGP attributes. The perturbation is applied by adding a random integer offset $\Delta \in [1, 4]$ to all node features in the corresponding attribute column, yielding a balanced 7-class dataset.

To evaluate adaptability under topology variations, we construct three datasets, with key parameters reported in Table~\ref{tab_dataset_configurations}. Specifically, the \textit{baseline} dataset contains randomly generated topologies with a moderate scale, used for training; the \textit{larger-scale} dataset simulates network expansion; and the \textit{real-world} dataset comprises operational ISP topologies from the Internet Topology Zoo~\cite{TopologyZoo}, a collection of actual network topologies from all over the world.

\subsection{Benchmark Algorithms}
We compare our EtaGATv2 against three baselines:
\begin{itemize}
    \item \textit{GAT}~\cite{GAT_ICLR_2018}, the foundational GAT that computes attention weights based on node features;
    \item \textit{GATv2}~\cite{GATv2}, an improved variant with the dynamic attention mechanism that addresses GAT's static attention limitation;
    \item \textit{EtaGAT}, which serves as an ablation study benchmark to isolate the contribution of edge-type awareness from the dynamic attention of our EtaGATv2 algorithm.
\end{itemize}
We train all models with batch size $4$, learning rate $10^{-4}$, $2$ attention layers, $8$ attention heads, hidden dimension $128$, and the Adam optimizer (weight decay $10^{-5}$) for $400$ epochs.

\begin{figure*}[t]
    \centering
    \includegraphics[width=18cm, height=4.8cm]{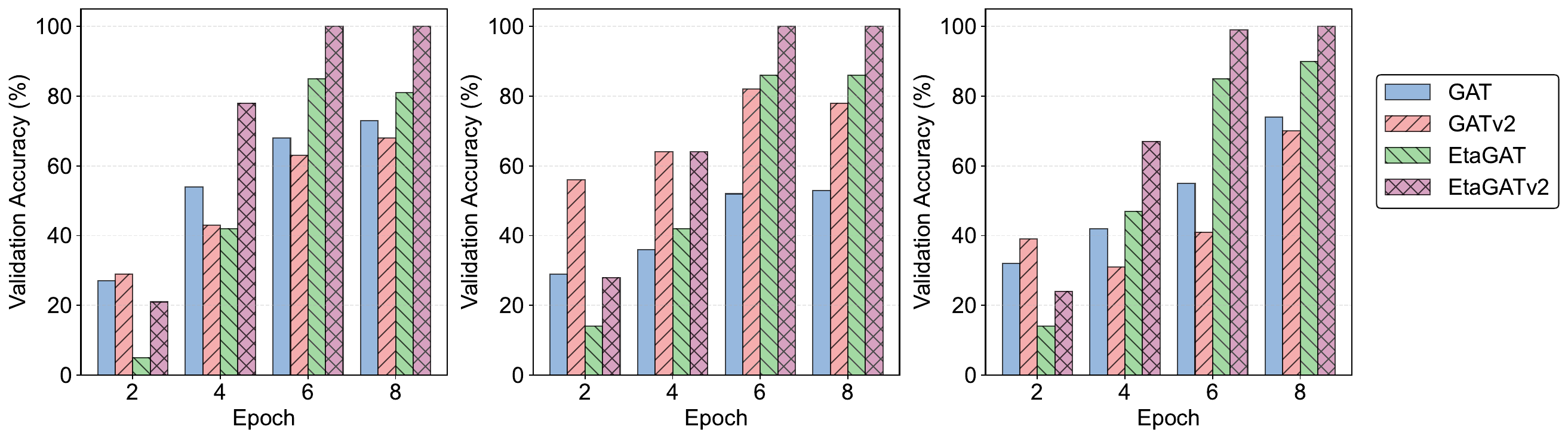}
    \caption{Validation across topology variations. From left to right are the baseline, larger-scale, and real-world datasets that are summarized in Table~\ref{tab_dataset_configurations}, with 100 samples in each validation epoch.}
    \label{fig_validation_comparison}
\end{figure*}

\subsection{Results and Analysis}
Fig.~\ref{fig_sim_training} shows a plot of the performance along the training epochs, with each curve corresponding to five independent runs. In particular, Fig.~\ref{fig1_training_loss_4algo} shows that both the proposed EtaGATv2 and the baseline algorithms converge with a steady training loss; while Fig.~\ref{fig2_training_accuracy_4algo_moving_avg} shows that EtaGATv2 outperforms the baselines in accuracy at any parity of training epochs, demonstrating higher training efficiency.
Fig.~\ref{fig2_training_accuracy_4algo_moving_avg} also validates that both the dynamic attention and edge-type-aware mechanisms help improve training efficiency in our misconfiguration classification task. We can observe from this figure that EtaGATv2 reaches 80\% accuracy at approximately $7\times10^3$ training samples, while GATv2 requires about $1.4\times10^4$ samples to achieve the same accuracy, demonstrating a reduction of roughly 50\% in required training data. This sample efficiency is critical for real-world networks where negative-labeled misconfiguration data are limited.

To evaluate adaptability under dynamic topologies, Fig.~\ref{fig_validation_comparison} reports the zero-shot validation accuracy where models trained for $\{2, 4, 6, 8\}$ epochs on the baseline dataset are tested on all three datasets. Across all topologies, EtaGATv2 achieves the highest accuracy for 4 epochs or more, confirming its superior adaptability to topology changes with limited misconfiguration training. This is critical in dynamic industrial networks. Interestingly, GATv2 provides a strong performance on larger-scale topologies, indicating that its dynamic attention mechanism provides better adaptation for dynamic topologies.

\section{Conclusion and Future Work}
In this work, we first mathematically formulated a protocol misconfiguration classification problem for template-based misconfigurations in networks. To solve this problem, we proposed our EtaGATv2 algorithm, a lightweight approach of linear complexity. Experimental results over three probing datasets have shown that EtaGATv2 has proved capable of greater sample efficiency, reducing required training samples by approximately 50\% compared to state-of-the-art baselines, making it suitable for wireless communication networks with dynamic topologies and limited negative-labeled data.
As future work, we plan to extend this initial study with misconfiguration models with more complicated operational error patterns, multiple concurrent misconfigurations, and explicit specification-configuration mappings.


\footnotesize
\bibliographystyle{IEEEtran}
\bibliography{IEEEabrv,refs.bib}




\end{document}